\documentclass[fleqn,12pt,twoside]{article}
\usepackage{espcrc1}

\usepackage{graphicx}
\usepackage[figuresright]{rotating}

\newcommand{\AmS}{{\protect\the\textfont2
  A\kern-.1667em\lower.5ex\hbox{M}\kern-.125emS}}

\title{Theoretical understanding of the nuclear incompressibility:
       where do we stand ?}

\author{G. Col\`o\address[MI]{Dipartimento di Fisica, Universit\`a degli Studi, 
                              \\and INFN sez. di Milano, via Celoria 16, 20133 
                              Milano (Italy)} 
        and
        Nguyen Van Giai\address{Institut de Physique Nucl\'eaire, IN2P3-CNRS,
        91406 Orsay Cedex (France)}
       }
       
\begin{document}

\maketitle

\begin{abstract}
The status of the theoretical research on the compressional modes 
of finite nuclei and the incompressibility $K_\infty$ of nuclear matter, 
is reviewed. It is argued that the recent experimental data on the 
Isoscalar Giant Monopole Resonance (ISGMR)
allow extracting the value of $K_\infty$ with an uncertainity of about 
$\pm$ 12 MeV.
Non-relativistic (Skyrme, Gogny) and relativistic mean field models
predict for $K_\infty$ values which are 
significantly different from one another, namely $\approx$ 220-235 and 
$\approx$ 250-270 MeV respectively. It is shown that the solution 
of this puzzle requires
a better determination of the symmetry energy at, and around, saturation. 
The role played by the experimental data of the Isoscalar Giant Dipole 
Resonance (ISGDR) is also discussed. 
\end{abstract}

\section{INTRODUCTION}
\label{intro}

The quest for the value of the nuclear incompressibility $K_\infty$ is still
continuing. Some significant progress in our understanding of how its value
can be constrained have been achieved in recent times and this 
will constitute the
subject of the present review. In this sense, this contribution is 
a continuation of those by J.P. Blaizot\cite{Blaizot:1998} and by N. Van Giai 
{\em et al.}\cite{NVG:2001} in the
previous conferences of the Giant Resonance series 
(Varenna 1998 and Osaka 2000).  

It is well known that the energy per particle $E/A$ in nuclear matter, 
considered as a 
function of the density $\varrho$, exhibits a minimum at the saturation 
point $\varrho_0$=0.17 fm$^{-3}$. The nuclear matter 
incompressibility, defined as 
\begin{equation}
K_\infty = 9\varrho_0^2 {d^2\over d\varrho^2} 
\left. {E\over A} \right|_{\varrho=\varrho_0},
\label{defK}
\end{equation}
provides a measure of the curvature of $E/A$ around $\varrho_0$. 
The interest of determining $K_\infty$ stems also from its impact
on the physics of neutron stars. 

Since we cannot directly create and probe nuclear matter in ordinary
laboratories, the only way to
extract a value for $K_\infty$ is by making contact with the
phenomenology of the compressional modes in finite nuclei. 
The clearest example of compressional mode is the Isoscalar Giant Monopole 
Resonance (ISGMR), which is often called the nuclear ``breathing mode''
and is excited by the operator
\begin{equation}
\hat M = \sum_{i=1}^{\rm A} r_i^2.
\label{opr_monopole}
\end{equation} 
The first evidences of this $L=0$ mode date back to the 1970s. The presence, 
in the same energy region, of modes with different multipolarities 
(e.g., $L=2$)
as well as of a non negligible background, makes the extraction of the 
monopole strength rather difficult. Over the last two decades, the experimental
techniques have improved and our knowledge of the ISGMR properties has
progressed. Many reactions have been employed to study this resonance, but
inelastic ($\alpha,\alpha'$) scattering has been, from the beginning, one of
the best tools. In the light nuclei this resonance is rather fragmented, while in
the medium-heavy nuclei it corresponds to a single peak of energy 
$E_{\rm ISGMR} \sim$ 80$\cdot$A$^{-1/3}$ MeV.
Nowadays, in the recent measurements of $^{90}$Zr, $^{116}$Sn, 
$^{144}$Sn and $^{208}$Pb performed at 
Texas A\&M~\cite{Youngbloodetal:1999}, the accuracy on the centroids of the 
ISGMR strength distribution has come down to about $\pm$2\%. The importance 
of this high accuracy for the extraction of $K_\infty$ will be discussed
below. 

\section{THE NUCLEAR INCOMPRESSIBILITY DEDUCED FROM THE ISGMR: 
BASIC FORMULAS AND PREVIOUS RESULTS}
\label{basic}

To perform the link between the properties of the ISGMR in finite nuclei and
$K_\infty$, the definition (\ref{defK}) must be complemented by an operative 
expression which contains quantities that can be measured in the 
laboratory. J.P. Blaizot~\cite{Blaizot:1980} showed that a plausible 
{\em definition} of the finite nucleus incompressibility $K_A$ is given by 
\begin{equation}
E_{\rm ISGMR} = \sqrt{\hbar^2 K_{\rm A}\over m<r^2>},
\label{KAdef}
\end{equation}
where $m$ is the nucleon mass, and $<r^2>$ the ground state mean square
radius. Using the experimental ISGMR energies to deduce $K_{\rm A}$ for 
different values of A, there have been attempts to make an extrapolation of
$K_{\rm A}$ to A=$\infty$. This was done by using a Weisz\"acker-type formula for
$K_{\rm A}$, namely
\begin{equation}
K_{\rm A} = 
K_\infty + K_{surf}{\rm A}^{-1/3} 
+ K_{sym}\alpha^2 + K_{Coul}{{\rm Z}^2\over {\rm A}^{4/3}},
\label{KAexp}
\end{equation}
where $\alpha={{\rm N-Z \over A}}$. M. Pearson~\cite{Pearson:1991} was the 
first to show that, in view of the correlations among the parameters of 
(\ref{KAexp}) and the scarcity of experimental data, trying to
use these to perform a fit of Eq. (\ref{KAexp}) is statistically meaningless 
and would indeed leave $K_\infty$ basically undetermined (fits which lead to 
100 MeV or 400 MeV may be equally acceptable). Similar conclusions 
were reached by S. Shlomo and D. Youngblood \cite{ShlomoYoungblood:1993}. 
Therefore, we will not discuss these so-called ``macroscopic approaches'' 
to $K_\infty$. However, we remark that the theoretical values of the 
parameters entering Eq. (\ref{KAexp}) can be calculated, within the 
framework of different models. We will come back to this point in 
Sec.~\ref{relativistic}.

The procedure to obtain $K_\infty$, which is nowadays believed to be 
physically sound, is the so-called ``microscopic approach''. The basic
idea consists in using energy functionals $E[\varrho]$ which allow
calculating nuclear matter and finite nuclei on the same footing. 

In the non-relativistic case, the starting point is a two-body 
effective nucleon-nucleon interaction $V_{\rm eff}$, whose parameters
are adjusted to reproduce experimental data in a small set of nuclei. 
$E$ is written as the expectation value of $H_{\rm eff}=T+V_{\rm eff}$ on a 
independent particle wave function (i.e., a Slater determinant).
In practice, the available functionals are based on the Skyrme and Gogny
interactions. 
 
In the relativistic models, nucleons are described as Dirac particles which 
interact by the exchange of effective $\sigma$, $\omega$ and $\rho$ mesons. 
In the limit of large meson masses, point coupling models are obtained. In 
some cases, the coupling constants are taken to be density-dependent. As in 
the cases of Skyrme and Gogny, the coupling constants are fitted. A specific 
model has been recently proposed~\cite{Finelli_tbp}, in which 
the main parameters are connected to more fundamental quantities, in particular
to the so-called QCD sum rules and to the (iterated) pion exchange. In all 
models, the no-sea approximation is made and the energy functional is written
in the Hartree (and not Hartree-Fock) form. 

In both non-relativistic and relativistic cases, the second derivative
of the energy functional can be calculated analytically for uniform nuclear
matter and the value of $K_\infty$ associated to a given parametrization is 
therefore given. In the case of finite nuclei, one calculates the 
monopole excitation using self-consistent linear response theory. The
system is perturbed with an (arbitrarily small) external field and
the small oscillations around the ground state are governed by the
residual force ${\delta^2 E\over \delta^2 \varrho}$. The theory is known
as self-consistent Random Phase Approximation (RPA) and is well described
in textbooks~\cite{RingSchuck:1980}. 

Then, the determination of $K_\infty$ proceeds as follows. 
\begin{itemize}
\item Using a set of different parametrizations (within a given class of 
energy functionals) characterized by different values of $K_\infty$, 
self-consistent RPA calculations of the ISGMR are performed in a given nucleus.
If the monopole strength has only one peak, $E_{\rm ISGMR}$ is well defined 
and Eqs. (\ref{KAdef}-\ref{KAexp}) suggest that a relation of the type 
$E_{\rm ISGMR}\sim\sqrt{K_\infty}$ can be expected. This can be verified 
empirically and indeed, relations of the type 
\begin{equation} 
E_{\rm ISGMR} = a\sqrt{K_\infty}+b
\label{linear}
\end{equation}
have been interpolated (see below). 
\item The experimental value of $E_{\rm ISGMR}$ is inserted in 
Eq. (\ref{linear}) and the value of $K_\infty$ is deduced. 
\end{itemize}

One or few nuclei would be enough to apply this procedure, and $^{208}$Pb 
is a typical system where the monopole strength has a well-defined peak and 
where recent experiments have reduced the sources of errors. From the 
theoretical point of view, the calculations must be free of the 
uncertainites associated to, e.g., the description of pairing or to 
anharmonic effects. 

The procedure that we have described, was firstly applied by J.P. Blaizot 
and collaborators~\cite{Blaizot:1995}, by employing the Gogny interaction. 
The authors of~\cite{Blaizot:1995} made use of the existing parametrizations 
and also built {\em ad hoc} new ones in order to cover more values 
of $K_\infty$. A value of about 230 MeV for the nuclear incompressibility 
can be extracted from the experimental $E_{\rm ISGMR}$ of $^{208}$Pb. 
Exactly the same procedure was applied in the case of the Skyrme forces 
(using only already existing parametrizations)~\cite{Colo:2001,NVG:2001}. 
From the experimental monopole energy of $^{208}$Pb a value of $K_\infty$ 
around 210 MeV is deduced, while using $^{90}$Zr the value is even lower 
(around 200 MeV). All the RPA calculations quoted are done using a discrete 
basis. However, the value of 210 MeV is consistent with what has been found 
by other authors within the framework of continuum-RPA \cite{Hamamoto:1997}. 

Different calculations made by using the relativistic RPA gave instead 
larger values of $K_\infty$. Values of 250 MeV and 270 MeV were extracted 
from the experimental data in $^{208}$Pb and $^{144}$Sm 
respectively~\cite{Ma:2001,Niksic:2002}. This model 
dependence in the extraction of $K_\infty$ has been, and partly it is
still, the basic puzzle. 

\section{CONSISTENCY BETWEEN THE RESULTS FROM SKYRME AND GOGNY INTERACTIONS}
\label{NR}

Before proceeding, some quantitative statements concerning the numerical 
accuracy of the microscopic approach are in order. From Eq. (\ref{linear}), 
the relative errors on the ISGMR energy and on
$K_\infty$ are related by
\begin{displaymath}
{\delta K_\infty\over K_\infty}=2{\delta E_{\rm ISMGR}\over E_{\rm ISGMR}}.
\end{displaymath}
As a rule of thumb, let us keep in mind that sticking to the case of 
$^{208}$Pb, $\pm$150 keV of uncertainity on $E_{\rm ISMGR}$ result in 
about $\pm$5 MeV uncertainity on $K_\infty$. The experimental measurement on
$^{208}$Pb   
provides us with 14.17 $\pm$ 0.28 MeV \cite{Youngbloodetal:1999}
and therefore 
\begin{displaymath}
\delta K_\infty^{\rm exp.} \sim \pm 10\ {\rm MeV}. 
\end{displaymath}
Theoretically, the best way to extract the monopole energy is by means of 
constrained Hartree-Fock (CHF) calculations~\cite{Bohigas:1979}. These
calculations provide the inverse energy-weighted sum rule $m_{-1}$ 
with a numerical error 
of the order of $\pm$3\% (see also Ref.~\cite{Meyer_tbp}), and
since the energy-weighted sum rule $m_1$ is known, the relation 
\begin{equation}
E_{\rm ISGMR} = \sqrt{m_1\over m_{-1}}
\label{defE} 
\end{equation}
implies
\begin{displaymath}
\delta K_\infty^{\rm th.} \sim \pm 7\ {\rm MeV}.
\end{displaymath}
The two errors on $K_\infty$ are independent and should be added 
quadratically, so that 
\begin{equation}
\delta K_\infty \sim \pm 12\ {\rm MeV}. 
\end{equation}

\begin{table}[htb]
\caption{Values of the $m_{-1}$ sum rule in fm$^4$/MeV calculated using
the interaction SLy4. In column (a), results obtained without the Coulomb
and spin-orbit interaction are shown. In this case, the RPA is fully 
self-consistent. In column (b), the Coulomb and spin-orbit terms of the
interaction are included in the CHF calculation and in the HF calculation
on which RPA is based, but are excluded from the residual RPA interaction.}
\label{table:1}
\newcommand{\m}{\hphantom{$-$}}
\newcommand{\cc}[1]{\multicolumn{1}{c}{#1}}
\renewcommand{\tabcolsep}{2pc} 
\renewcommand{\arraystretch}{1.2} 
\begin{tabular}{@{}llll}
          &     & (a)          & (b)          \\
\hline
$^{16}$O  & CHF & \cc{$14.45$} & \cc{$16.04$} \\
          & RPA & \cc{$14.49$} & \cc{$16.73$} \\
\hline
$^{40}$Ca & CHF & \cc{$75.31$} & \cc{$88.31$} \\
          & RPA & \cc{$75.91$} & \cc{$92.13$} \\
\hline
\end{tabular}\\[2pt]
\end{table}

\begin{figure}[htb]
\begin{tabular}{@{}l}
\includegraphics[scale=1.2]{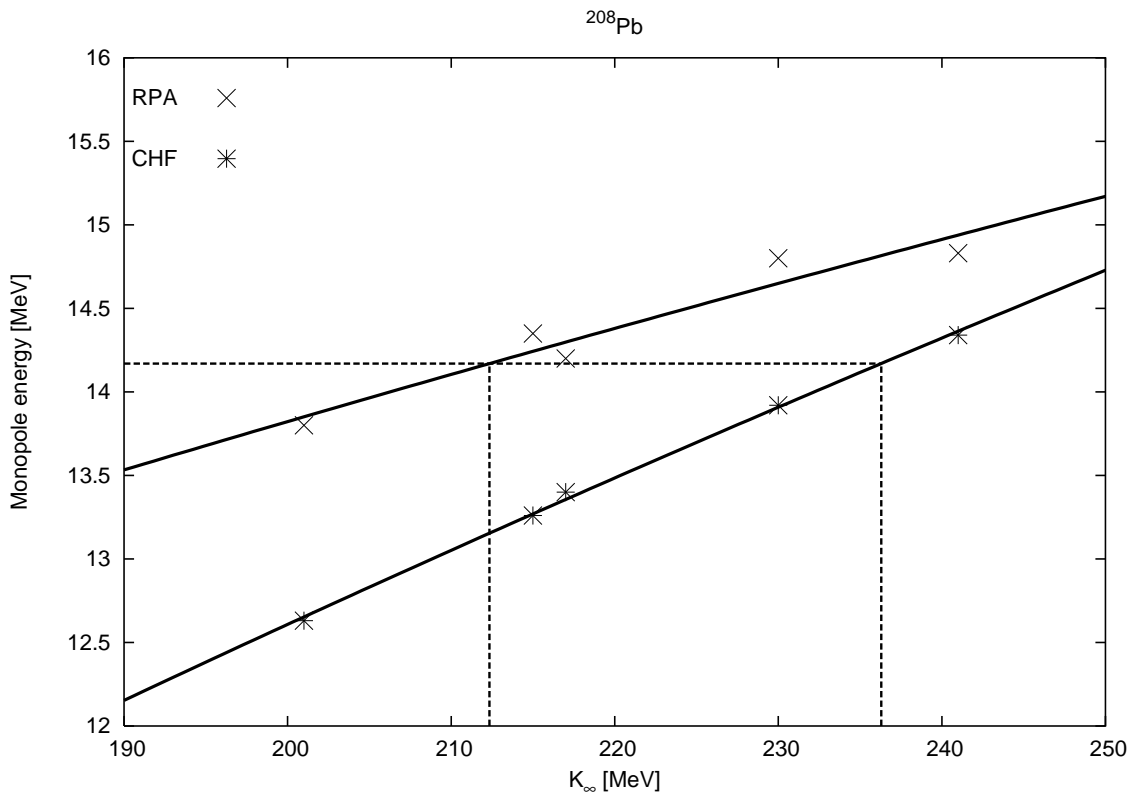}\\
\end{tabular}
\caption{Monopole energies in $^{208}$Pb (defined as in Eq.~(\ref{defE})) 
obtained from RPA (crosses) and CHF (stars) calculations which employ 
different Skyrme forces, plotted as a function of the associated $K_\infty$. 
The lines are fits of the type (\ref{linear}). The dashed lines indicate
the extracted values of $K_\infty$. The value of $K_\infty$ resulting from 
CHF is in agreement with that extracted from the Gogny calculations 
of~\cite{Blaizot:1995}.}
\label{fig:1}
\end{figure}

\begin{figure}[htb]
\begin{tabular}{@{}l}
\includegraphics[scale=1.2]{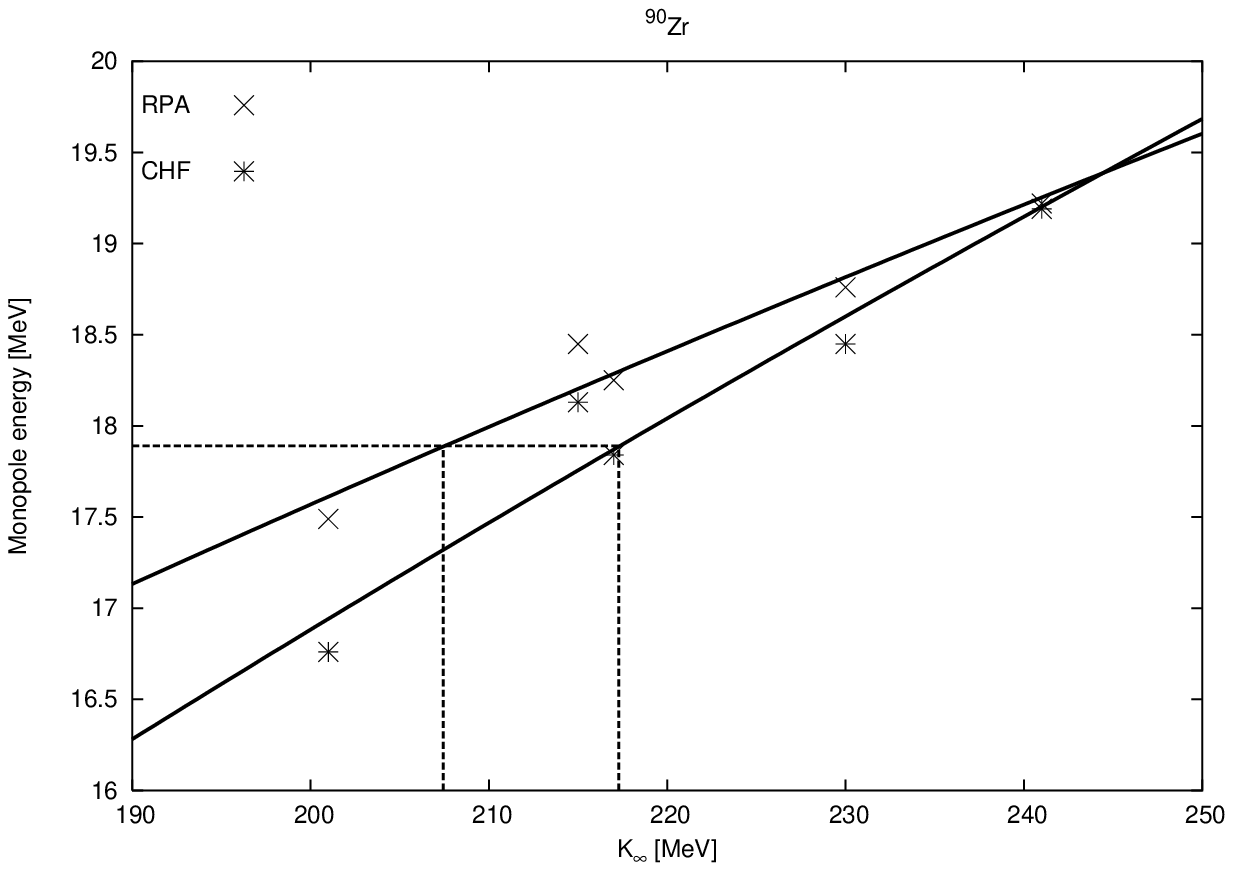}\\
\end{tabular}
\caption{The same as Fig.~\ref{fig:1} in the case of $^{90}$Zr.}
\label{fig:2}
\end{figure}

We now discuss in more detail the calculations performed in 
Refs.~\cite{Colo:2001,NVG:2001}. One of their main limitations is the lack of 
complete self-consistency, as the residual p-h Coulomb 
and p-h spin-orbit interactions 
are dropped. We have analyzed the effect of this approximation in the 
nuclei $^{16}$O and $^{40}$Ca. The results are shown in Table~\ref{table:1}. 
The Skyrme force used is SLy4~\cite{Chabanat:1998}. The results of (a)
are obtained by dropping the Coulomb and spin-orbit terms both in HF and in 
RPA. In this sense, the calculation is fully self-consistent, and the fact 
that the RPA results agree with the CHF results within less than 1\% suggests 
that the approximations done in the RPA (neglect of the continuum and truncation
of the discrete basis) do not affect seriously the values of $m_{-1}$. On the
other hand, the results of (b) are obtained by including the Coulomb and 
spin-orbit terms in the mean field and not in the residual RPA interaction. 
The difference between the CHF and RPA results is larger than the CHF 
intrinsic uncertainity and can therefore be a meaningful indication of the error 
induced by the lack of full RPA self-consistency. Consequently, we have compared 
CHF and RPA in the nuclei already considered for the extraction of $K_\infty$,
that is, $^{208}$Pb and $^{90}$Zr. Note that in the case of $^{208}$Pb and $^{90}$Zr 
the CHF result for $m_{-1}$ is larger than the RPA result, contrarily to the 
case of $^{16}$O and $^{40}$Ca. The results are displayed in Figs.~\ref{fig:1} 
and~\ref{fig:2}. 

%
%

The values of 235 MeV from $^{208}$Pb and 220 MeV from $^{90}$Zr for 
$K_\infty$ which are extracted from the Skyrme-CHF calculations, are
consistent with each other and with the values deduced using the Gogny 
interaction. This is the first important conclusion of the present paper. 

We conclude this Section with a brief justification of the mean field 
approximation in the present calculations. Firstly, it has been shown in
Ref.~\cite{Colo:1992} that including the coupling of the RPA states with
more complicated configurations of 2 particle-2 hole (2p-2h) type, shifts 
the ISGMR in $^{208}$Pb to lower energy by only $\approx$ 500 keV. This
number is smaller than the uncertainities discussed above. Secondly, 
according to the arguments of Sec.~\ref{basic}, in order to use the
result for $E_{ISGMR}$ obtained beyond mean field, a calculation for
nuclear matter performed on the same footing is needed. The development
of a suitable energy functional is still a challenge for nuclear structure
theory.  

\section{HOW TO RECONCILE THE RELATIVISTIC MEAN FIELD WITH SKYRME 
AND GOGNY ?}
\label{relativistic}

As far as the relativistic models are concerned, there has been a recent
suggestion~\cite{Piekarewicz:2003} that the different outcome for $K_\infty$
(compared with Skyrme and Gogny) is originated by
the different density dependence of the symmetry energy $S(\rho)$ predicted by
different models.
It is true that the
measurements of the ISGMR are done in a system ($^{208}$Pb) with a finite
value of ${\rm (N-Z)/A}$.
To illustrate the possible influence of the density dependence of $S(\rho)$
on the extraction of $K_\infty$, J. Piekarewicz~\cite{Piekarewicz:2003} has
built parametrizations of effective Lagrangians whose symmetry energy has
different density dependences (this is easy to achieve since the $\rho$
coupling constant is an adjustable parameter) and he finds that the
extracted $K_\infty$ indeed differ and can even become close to Skyrme force
values. To achieve this, it is necessary to soften  the function $S(\rho)$
and in Ref.~\cite{Piekarewicz:2003} this was done by lowering the symmetry
energy at the saturation point, $a_\tau$. However, it has been pointed out
in Ref.~\cite{Vretenar:2003} that, in the type of model used in
Ref.~\cite{Piekarewicz:2003} parametrizations with $a_\tau$ lower than 36
MeV cannot describe satisfactorily $N \ne Z$ nuclei.   
A complementary attempt has been made 
in~\cite{Meyer_tbp}, by constructing Skyrme forces with associated values
of $a_\tau$ up to 38 MeV. In this case, finite nuclei have been 
carefully considered in the fits. By calculating the ISGMR centroid energy, 
a very weak dependence on $a_\tau$ has been found. 

\begin{figure}[htb]
\begin{tabular}{@{}l}
\includegraphics[scale=1.2]{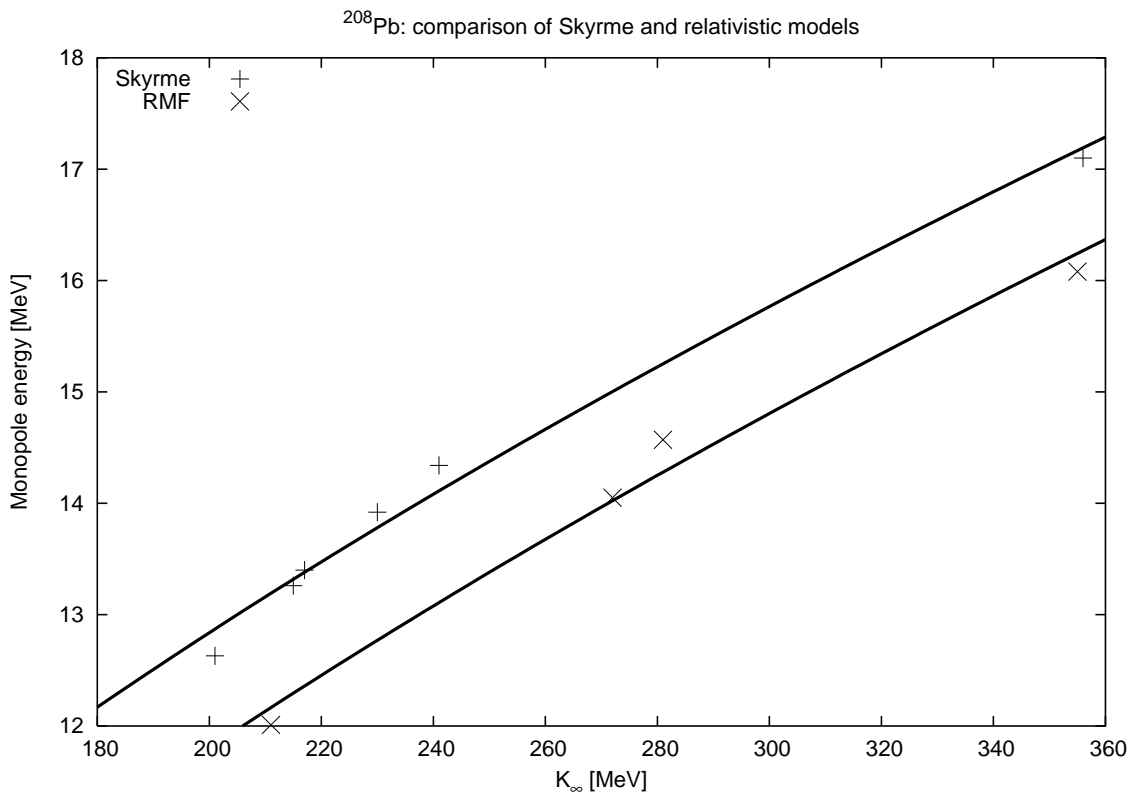}\\
\end{tabular}
\caption{Monopole energies as a function of $K_\infty$ in Skyrme and in
relativistic mean field (RMF) models. The relativistic results are taken 
from Ref.~\cite{Ma:2001}. The Skyrme results are the same as in 
Fig.~\ref{fig:1} (only the point corresponding to the SIII interaction 
has been added in order to compare with the relativistic result for 
$K_\infty\approx$ 350 MeV). The lines are numerical fits.}
\label{fig:3}
\end{figure}

There is one more conceptual remark. Starting from (\ref{KAexp}), one would 
expect a dependence of $K_{\rm A}$, and consequently of the monopole energy, 
on a parameter like $K_{sym}$ more than on $a_\tau$. 
That is,
on the derivatives of the symmetry energy more than on its value 
at saturation. In fact, it has been found 
that both in relativistic and in non-relativistic 
models the quantity $K_{surf}$ is essentially given by $c K_\infty$ with 
$c\approx$ -1 \cite{Blaizot:1980,Centelles:2002}. Therefore, we can write
\begin{eqnarray}
K_{\rm A} & \sim & 
K_\infty^{({\rm non\ rel.})}  (1+c{\rm A}^{-1/3}) + K_{sym}^{({\rm non\ rel.})}\alpha^2 
+ K_{Coul}^{({\rm non\ rel.})} {{\rm Z}^2\over {\rm A}^{4/3}},
\nonumber \\
K_{\rm A} & \sim & 
K_\infty^{({\rm rel.})} (1+c{\rm A}^{-1/3}) + K_{sym}^{({\rm rel.})}\alpha^2 
+ K_{Coul}^{({\rm rel.})} {{\rm Z}^2\over {\rm A}^{4/3}}.
\label{KAcompare}
\end{eqnarray}
It is likely that the third term of the r.h.s. (Coulomb contribution) 
does not change much from a non-relativistic to a relativistic description.
The same values of $K_{\rm A}$ can thus be obtained with different values 
of the $K_\infty$ and $K_{sym}$ terms. 
For illustration, we display in Fig.~\ref{fig:3} the results of
the Skyrme calculations of the ISGMR in $^{208}$Pb (already shown in
Fig. 1) together with the corresponding results of the relativistic mean field
taken from\ \cite{Ma:2001}. In this nucleus we have $\alpha^2=0.04$. 
At any given value of $K_\infty$ the difference between the two curves 
is approximately 1 MeV. This translates into a difference of about 20 
MeV in $K_{\rm A}$. If this difference is entirely attributed 
to the negative term $K_{sym} \alpha^2$ of Eq. (\ref{KAcompare}), the 
values of $K_{sym}$ in the non-relativistic and relativistic models
would differ by about 500 MeV. Only few calculations of
$K_{sym}$ are available\  \cite{Blaizot:1980,Centelles:2002}, 
therefore more systematic tests of the present argument should be made.
More importantly, we apparently miss any experimental constraint
to decide what is the proper value of the parameter $K_{sym}$. 

\section{THE ISOSCALAR GIANT DIPOLE RESONANCE}
\label{ISGDR}

The ISGDR is a non-isotropic compressional mode, which is excited by
the operator 
\begin{equation}
\hat D = \sum_{i=1}^{\rm A} r_i^3 Y_{1M}(\hat r_i),
\label{opr_dipole}
\end{equation} 
and which provides in principle a further way to extract the value of 
$K_\infty$. There are many reasons 
why this task is more 
challenging in the case of the dipole than in the case of the monopole. 

From the theoretical point of view, one difficulty arises from the fact 
that the spurious center-of-mass translation (associated with the 
operator $\sum_{i=1}^{\rm A} r_i Y_{1M}(\hat r_i)$) carries the same
quantum numbers as the ISGDR. In principle, a sharp spurious state 
at zero energy should result from an ideal self-consistent RPA  
calculation. However, in practice this is not the case due to
approximations and numerical 
inaccuracy. Consequently, the resulting states
are not orthogonal to the spurious state and one has to correct for this.
The spurious transition density is expected to be of the type 
$\sim {d\varrho_0\over dr}$ where $\varrho_0$ is the ground state density. 
It is possible to project out the spurious component from each excited
state (cf., e.g., Ref.~\cite{Colo:2000}). An equivalent procedure~\cite{NVG:1981}
consists in using instead the modified operator
\begin{displaymath}
\hat D_{eff} = \sum_{i=1}^{\rm A} (r_i^3-\eta r_i) Y_{1M}(\hat r_i),
\end{displaymath} 
where $\eta={5\over 3}<r^2>$. A recent analysis of the accuracy of these
techniques, as well as a discussion about some different prescriptions to
subtract the spurious state, can be found in Ref.~\cite{Hamamoto:2002}. There
have been other discussions in the recent literature~\cite{Shlomo:2002}, also
in connection with semiclassical models~\cite{Abrosimov:2002}.

On the experimental side, different ($\alpha,\alpha'$) measurements have 
been performed over the years but the problem of disentangling 
the ISGDR strength from the other multipoles (and from the IVGDR) is 
far from being trivial. The ISGDR lies at higher 
energies than the ISGMR (approximately 110$\cdot$A$^{-1/3}$ MeV). An accurate 
determination of its high-energy tail is therefore more difficult. On the 
low-energy side, a sizeable amount of fragmented strength is found. These issues 
are discussed in these Conference proceedings~\cite{Umesh_here,Lui_here}.

\begin{figure}[htb]
\begin{tabular}{@{}l}
\includegraphics[scale=0.5]{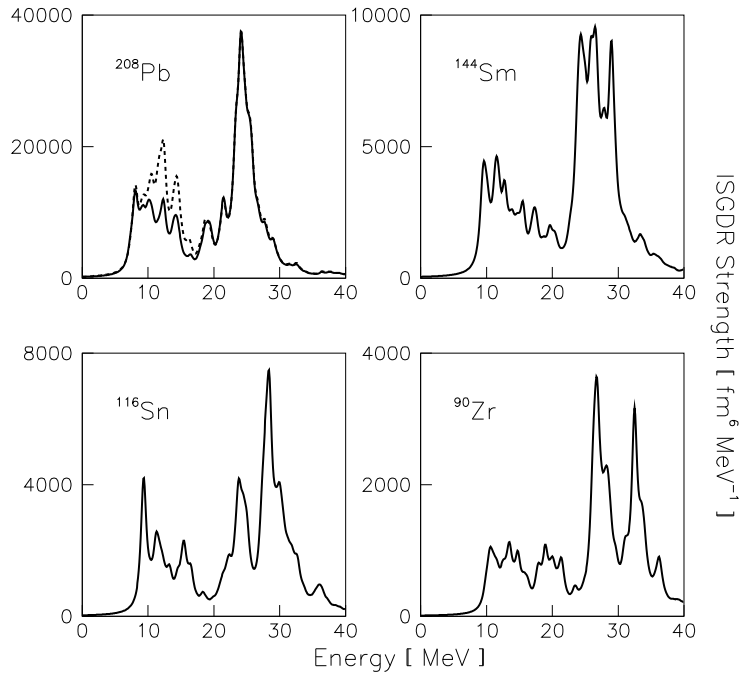}\\
\end{tabular}
\caption{ISGDR strength functions in different nuclei, calculated in RPA
using the Skyrme interaction SGII~\cite{NVG:1981b} and corrected for 
center-of-mass effects. In the case of $^{208}$Pb the dashed line corresponds 
to a calculation where the spurious center-of-mass state is not subtracted. 
Here as well as in Fig. 5 the discrete RPA states have been smeared
out by means of a Lorentzian of 1 MeV width. Taken from Ref.~\cite{Colo:2000}.}
\label{fig:4}
\end{figure}

Different theoretical calculations~\cite{Colo:2000,Vretenar:2000}, 
have clarified that the low-energy part of the ISGDR strength is
formed by non-collective states. The separation of a ``high-energy'' region 
and a ``low-energy'' region in the ISGDR strength distribution, emerges
systematically from the calculations of Ref.~\cite{Colo:2000} 
as illustrated here in 
Fig.~\ref{fig:4}. One of the main indications about the different character 
of the two parts of the strength, comes from the fact that the centroids of 
the high-energy regions, calculated with different Skyrme forces in a given
nucleus, scale with the corresponding $K_\infty$ (which testifies to their 
compressional nature), whereas the centroids of the low-energy regions do not. 
The same pattern is found in the relativistic calculations. 

More detailed considerations concerning the wave functions of the low-lying 
states have been done in Ref.~\cite{Vretenar:2002}
where the authors
suggest that the low-lying strength may correspond to a ``toroidal'' 
resonance, which can be visualized - in a simple way - by thinking of
a particle current bent into a torus. It is excited by the vector operator
\begin{equation}
\hat T = \sum_{i=1}^{\rm A} \vec\nabla\times (\vec r_i\times\vec\nabla)
r_i^3 Y_{1M}(\hat r_i)
\label{opr_toroidal} 
\end{equation}
which couples to the transition current. To verify the toroidal nature of
the states, some other probe than the $\alpha$-particles should be tried, 
and ($e,e'$) experiments would be probably useful in this respect.  

In Table~\ref{table:2} we report the results of the various self-consistent
RPA calculations of the ISGDR in $^{208}$Pb. The first remark is that the
low-energy part is of course significantly dependent on the model and on the
specific functional used, as expected due to its lack of collectivity. 
As far as the high-lying centroid is concerned, the Skyrme results of
Refs.~\cite{Hamamoto:1998,Colo:2000} (continuum and discrete RPA respectively)
are in good agreement with each other. In Ref.~\cite{Colo:1999} it has been
shown that coupling the RPA states with 2p-2h type configurations, 
shifts the ISGDR centroid down to 22.9 MeV, in very good agreement with experiment
(this coupling also produces a conspicous spreading width of about 6 MeV).

\begin{table}[htb]
\caption{Self-consistent (relativistic and non-relativistic) RPA calculations
performed for the ISGDR in $^{208}$Pb compared with the most recent experimental 
data. The two columns report the centroid energies (in MeV) of the low-energy 
and high-energy regions discussed in the text.}
\label{table:2}
\newcommand{\m}{\hphantom{$-$}}
\newcommand{\cc}[1]{\multicolumn{1}{c}{#1}}
\renewcommand{\tabcolsep}{2pc} 
\renewcommand{\arraystretch}{1.2} 
\begin{tabular}{@{}lll}
          & High-energy  & Low-energy          \\
          & centroid     & centroid            \\
\hline
Hamamoto {\em et al.}~\cite{Hamamoto:1998}  & \cc{$23.4$} & \cc{$\sim 14$} \\
Col\`o {\em et al.}~\cite{Colo:2000}        & \cc{$23.9$} & \cc{$10.9$} \\
Vretenar {\em et al.}~\cite{Vretenar:2000}  & \cc{$26$}   & \cc{$10.4$} \\
Piekarewicz~\cite{Piekarewicz:2001}         & \cc{$24.4$} & \cc{$\sim 8$} \\
Shlomo and Sanzhur~\cite{Shlomo:2002}       & \cc{$\sim 25$} & \cc{$\sim 15$} \\
\hline
Uchida {\em et al.}~\cite{Uchida:2003,Umesh_here} 
                                     &\cc{$23\pm 0.3$}  &\cc{$12.7\pm 0.2$} \\
Lui {\em et al.}~\cite{Lui_here}     &\cc{$21.7$}       &\cc{$12.6$} \\
\hline
\end{tabular}\\[2pt]
\end{table}

In Table 2, also the results of the most recent measurement performed at 
RCNP in Osaka (making use of $\alpha$-particles at incident energy of 400 MeV)
as well as the new findings of the Texas A\& M group, are shown. 
The experimental 
energies seem to converge towards each other 
(compared to the previous experiments of
the same groups, where the discrepancies were larger). Nevertheless, 
a point should be made concerning the experimental analysis. 
The directly measured 
quantity is the double-differential cross section, 
${d^2\sigma\over d\Omega dE}$. 
The multipole decomposition of this cross section is 
done by relying on Distorted 
Wave Born Approximation (DWBA) calculations where the radial form factors for 
the various multipole excitations are the same at all energies. 
This is somehow 
in contrast with the outcome of theory, as we have discussed above. 

We conclude the discussion on $^{208}$Pb with a statement about the deduction of
$K_\infty$. If this quantity is derived from a plot similar to those of 
Figs.~\ref{fig:1} and~\ref{fig:2}, i.e., by using the ``microscopic approach'' 
described in Sec.~\ref{basic} in connection with the ISGMR but using now the ISGDR data, 
one finds a value around 205 MeV which is still (marginally) 
compatible with the value 
220-235 MeV discussed above (cf. Fig. 2 of~\cite{Colo:2000}). This means that 
there is no basic contradiction between the present ISGDR and ISGMR data, as far as
the deduction of the nuclear incompressibility is concerned. However, it should be clear
from all our discussion, that we still need to await for improvements in the ISGDR
studies in order to reach the same confidence that we have in $K_\infty$ extracted 
from the ISGMR.

\begin{figure}[htb]
\begin{tabular}{@{}l}
\includegraphics[scale=1.2]{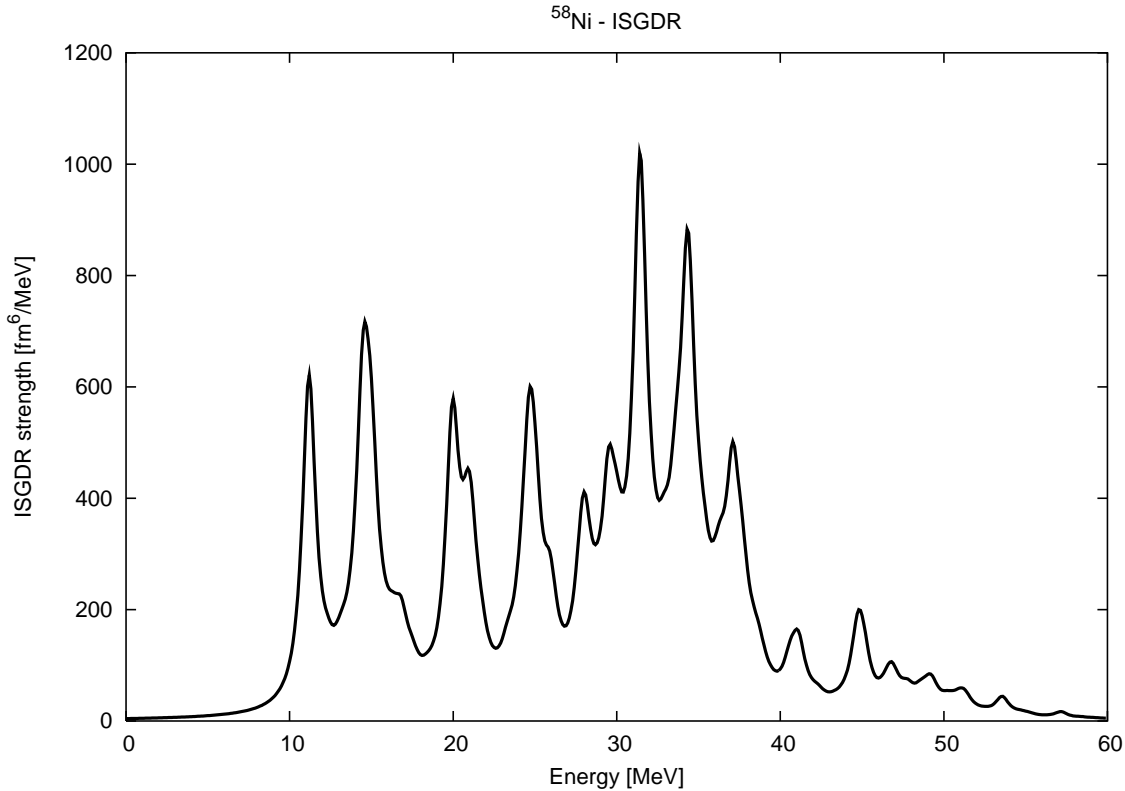}\\
\end{tabular}
\caption{ISGDR Strength distribution in $^{58}$Ni calculated using Skyrme-RPA with
the SGII force.}
\label{fig:5}
\end{figure}

We end this Section by showing the result of a Skyrme-RPA calculation for a lighter
system, namely $^{58}$Ni (see Fig.~\ref{fig:5}). It can be noticed, by comparison
with Fig.~\ref{fig:4}, that although the centroid of the ISGDR is still in the
region around 110$\cdot$A$^{-1/3}$ MeV, the strength is more fragmented than in
the heavier systems. Also, the distinction between a low- and a high-energy regions
is less evident. A comparison with experiment would be useful. In any case, as
it was said for the monopole case, this picture confirms that medium-mass systems
are less suitable for studying the nuclear incompressibility. 

\section{CONCLUSIONS}
\label{conclu}

In recent years, there have been significant progresses both in the 
experimental techniques aimed to extract with good precision the moments of the 
monopole strength function, and in the theoretical models - especially in those 
based on relativistic functionals which can be nowadays discussed on the same
footing as the well tested Skyrme and Gogny functionals. Still, it is shown in 
this paper that it is in general possible to determine a delicate
quantity like the nuclear incompressibility $K_\infty$ only within 
$\pm$ 12 MeV. 

Many discussions have been devoted to the fact that the values of $K_\infty$
depend on the model through which they have been extracted, either Skyrme,
Gogny or relativistic functionals. In this contribution, we show that the
discrepancy between Skyrme and Gogny does not exist and, using the ISGMR
data in $^{208}$Pb the value of $K_\infty$ lies between 220 MeV and 235 MeV. 
On the other hand, the relativistic calculations point to larger values, of 
the order of 250-270 MeV. This puzzle is still unsolved, but we argue in this
work that the reason has to be found probably in the different features that the 
asymmetry energy curve has in the relativistic and non-relativistic models,
respectively.  
The relativistic models are characterized by significantly larger values of the
symmetry energy at saturation, and of its first and second derivatives. 
The quest about the proper values of the asymmetry energy has of course very
general implications on the nuclear phenomenology, and new works on the subject
are in progress.

As far as the isoscalar dipole is concerned, there has been also definite
experimental improvements in the techniques aimed to reduce or eliminate the 
background, and in the data analysis. While in the past the results from 
different groups were in disagreement among themselves, and with theory, this
does not seem to be the case for the new experiments. The 
degree of accuracy that can be expected in extracting $K_\infty$ is smaller
in the dipole than in the monopole case. However, there does not seem to exist
a basic incompatibility between the values of the incompressibility deduced using
for instance the Skyrme forces, in the monopole and in the dipole case. Among
the perspectives, it can be said that new ($\alpha,\alpha'$) cross section calculations 
based on a fully microscopic input may be of great help to reduce some of the
uncertainities which still exist. On the other hand, it is unlikely that the ISGDR
data can help in solving the discrepancy between the non-relativistic and relativistic
models as far as the value of $K_\infty$ is concerned. 

\section*{ACNOWLEWDGMENTS}

Part of the work on the ISGMR stems from a collaboration with K. Bennaceur
and J. Meyer, whereas the ISGDR results were obtained in collaboration with 
P.F. Bortignon and M.R. Quaglia. We are also pleased to acknowledge 
stimulating discussions with J.P. Blaizot, M. Centelles, H. Clark, U. Garg, 
I. Hamamoto, Y.-W. Lui, J. Piekarewicz, P. Ring, H. Sagawa, H. Sakaguchi, 
S. Shlomo, M. Uchida, D. Vretenar, and D. Youngblood.


\begin{thebibliography}{9}
\bibitem{Blaizot:1998} J.P. Blaizot, Nucl. Phys. A649 (1999) 61c. 
\bibitem{NVG:2001} N. Van Giai, P.F. Bortignon, G. Col\`o, Z.-Y. Ma and
                   M.R. Quaglia, Nucl. Phys. A687 (2001) 44c. 
\bibitem{Youngbloodetal:1999} D. Youngblood, H.L. Clark and Y.-W. Lui, 
                              Phys. Rev. Lett. 82 (1999) 691. 
\bibitem{Blaizot:1980} J.P. Blaizot, Phys. Rep. 64 (1980) 171.
\bibitem{Pearson:1991} M. Pearson, Phys. Lett. B271 (1991) 12.
\bibitem{ShlomoYoungblood:1993} S. Shlomo and D. Youngblood, 
                                Phys. Rev. C47 (1993) 529.
\bibitem{Finelli_tbp} P. Finelli, N. Kaiser, D. Vretenar, W. Weise,
                      nucl-th/0307069. 
\bibitem{RingSchuck:1980} P. Ring and P. Schuck, The Nuclear Many-Body 
                          Problem, Springer-Verlag, New York, 1980.
\bibitem{Blaizot:1995} J.P. Blaizot, J.F. Berger, J. Decharg\'e, M. Girod, 
                       Nucl. Phys. A591 (1995) 435.
\bibitem{Colo:2001} G. Col\`o, N. Van Giai, P.F. Bortignon and M.R. 
                    Quaglia, in Proc. of the Conference ``Structure of the 
                    Nucleus at the Dawn of the Century'', Bologna 2000, 
                    eds. G. Bonsignori, M. Bruno, A. Ventura, D. Vretenar, 
                    World Scientific, 2001, p. 418.
\bibitem{Hamamoto:1997} I. Hamamoto, H. Sagawa and X.Z. Zhang, Phys. Rev. 
                        C56 (1997) 3121.
\bibitem{Ma:2001} Z.-Y. Ma, N. Van Giai, A. Wandelt, D. Vretenar, P. Ring, 
                  Nucl. Phys. A686 (2001) 173.
\bibitem{Niksic:2002} T. Nik\v{s}i\`{c}, D. Vretenar, P. Ring, 
                      Phys. Rev. C66 (2002) 064302.
\bibitem{Bohigas:1979} O. Bohigas, A.M. Lane, J. Martorell, 
                       Phys. Rep. 52 (1979) 267. 
\bibitem{Meyer_tbp} K. Bennaceur, G. Col\`o, J. Meyer and N. van Giai
                    (to be published).
\bibitem{Chabanat:1998} E. Chabanat, P. Bonche, P. Haensel, J. Meyer, 
                        R. Schaeffer, Nucl. Phys. A635 (1998) 231.  
\bibitem{Colo:1992} G. Col\`o, P.F. Bortignon, N. Van Giai, A. Bracco, 
                    R.A. Broglia, Phys. Lett. B276 (1992) 279.
\bibitem{Piekarewicz:2003} J. Piekarewicz, Phys. Rev. C66 (2003) 034305. 
\bibitem{Vretenar:2003} D. Vretenar, T. Nik\v{s}i\`{c}, P. Ring, 
                        Phys. Rev. C (in press).
\bibitem{Centelles:2002} S.K. Patra, M. Centelles, X. Vi\~{n}as and
                         M. Del Estal, Phys. Rev. C65 (2002) 044304. 
\bibitem{Colo:2000} G. Col\`o, N. Van Giai, P.F. Bortignon and M.R. Quaglia, 
                    Phys. Lett. B485 (2000) 362.
\bibitem{NVG:1981} N. Van Giai and H. Sagawa, Nucl. Phys. A371 (1981) 1. 
\bibitem{Hamamoto:2002} I. Hamamoto and H. Sagawa, Phys. Rev. C66 (2002) 
                        044315.
\bibitem{Shlomo:2002} S. Shlomo and A.I. Sanzhur, Phys. Rev. C65 (2002) 044310.
\bibitem{Abrosimov:2002} V.I. Abrosimov, A. Dellafiore, F. Matera, Nucl.
                         Phys. A697 (2002) 748.
\bibitem{Umesh_here} See the contribution by U. Garg in this volume.
\bibitem{Lui_here} See the contribution by Y.-W. Lui in this volume.
\bibitem{Vretenar:2000}  D. Vretenar, A. Wandelt, P. Ring, Phys. Lett.
                         B487 (2000) 334. 
\bibitem{NVG:1981b} N. Van Giai and H. Sagawa, Phys. Lett. B106 (1981) 379. 
\bibitem{Vretenar:2002} D. Vretenar, N. Paar, P. Ring and T. Nik\v{s}i\`{c}, 
                        Phys. Rev. C65 (2002) 021301.
\bibitem{Hamamoto:1998} I. Hamamoto, H. Sagawa and X.Z. Zhang, Phys. Rev.
                        C57 (1998) R1064.
\bibitem{Piekarewicz:2001} J. Piekarewicz, Phys. Rev. C64 (2001) 024307.
\bibitem{Uchida:2003} M. Uchida {\em et al.}, Phys. Lett. B557 (2003) 12.
\bibitem{Colo:1999} G. Col\`o, N. Van Giai, P.F. Bortignon and M.R. Quaglia, 
                    in Proc. of the the Conference `RIKEN Symposium and
                    Workshop on Selected Topics in Nuclear Collective
                    Excitations'', RIKEN, Wako city 1999, eds. Nguyen Dinh
                    Dang, U. Garg and S. Yamaji, RIKEN Review 23 (1999) 39.
\end{thebibliography}
\end{document}